\documentclass[12pt]{article}
\usepackage{hyperref}
\usepackage{graphicx}
\usepackage{graphics}
\usepackage{dsfont}
\usepackage{epsfig}
\usepackage{bbm}
\usepackage{amsmath,amssymb,amsthm,amscd}
\usepackage{float}
\usepackage{braket}

\usepackage{tikz}
\usetikzlibrary{shapes,arrows}
\usepackage{fancybox}
\usepackage[OT2,T1]{fontenc}
\DeclareSymbolFont{cyrletters}{OT2}{wncyr}{m}{n}
\DeclareMathSymbol{\Sha}{\mathalpha}{cyrletters}{"58}

\restylefloat{table}

\newcommand{\nn}{{\nonumber}}

\def\Mgn[#1]#2{{\overline{\cal M}_{#1,#2}}}
\def\pqs[#1,#2]{{\footnotesize{$\left[\begin{array}{c} #1\\#2  \end{array}\right]$}}} 
\def\pqsu[#1,#2]{\left[\begin{array}{c} #1\\#2  \end{array}\right]} 
\def\pqssu[#1,#2]{{\footnotesize{\left[\begin{array}{c} #1\\#2  \end{array}\right]}}} 
\def\pqh[#1,#2]{{\footnotesize{$\left[\begin{array}{c} #1\\#2  \end{array}\right]$}}} 
\def\pqhu[#1,#2]{\left[\begin{array}{c} #1\\#2  \end{array}\right]}

\newcommand{\ba}{\begin{eqnarray*}}
\newcommand{\ea}{\end{eqnarray*}}
\newcommand{\ban}{\begin{eqnarray}}
\newcommand{\ean}{\end{eqnarray}}
\newcommand{\be}{\begin{equation}}
\newcommand{\ee}{\end{equation}}
\newcommand{\ben}{\begin{equation}}
\newcommand{\een}{\end{equation}}
\numberwithin{equation}{section}

\newcommand{\IZ}{\mathbb{Z}}

\numberwithin{equation}{section}

\usepackage{hyperref}
\usepackage{amsmath}
\usepackage{amsthm}
\usepackage{amsfonts}          
\usepackage{amssymb}           
\usepackage[all]{xy}

\newcommand{\eqdef}{%
  \mathrel{\lower.1mm
    \hbox{$\stackrel{\lower.424ex\hbox{\scriptsize def}}{=}$}}
}

\newcommand{\R}{\ensuremath{\mathbb{R}}}
\newcommand{\C}{\ensuremath{\mathbb{C}}}
\newcommand{\Z}{\ensuremath{\mathbb{Z}}}

\newcommand{\tors}{\ensuremath{\text{tors}}}

\begin{document}
\begin{titlepage}
\begin{flushright}
\parbox[t]{1.73in}{\flushright 
UPR-1285-T}
\end{flushright}
\begin{center}

\vspace*{ 0.9cm}

{\large \bf {Type II String Theory on  Calabi-Yau Manifolds with Torsion\break\break and Non-Abelian Discrete Gauge Symmetries }}

\vspace*{ .2cm}
\vskip 0.8cm

\renewcommand{\thefootnote}{}
\begin{center}
 {Volker Braun$^1$, Mirjam Cveti\v{c}$^{2,3,4}$,   Ron Donagi$^{3,2}$, Maximilian Poretschkin$^2$}
\end{center}
\vskip .2cm
\renewcommand{\thefootnote}{\arabic{footnote}}

$\,^1$ {Elsenstrasse 35,
12435 Berlin,
Germany}\\[.3cm]
$\,^2$ {Department of Physics and Astronomy,\\
University of Pennsylvania, Philadelphia, PA 19104-6396, USA} \\[.3cm]
$\,^3$ {Department of Mathematics,\\
University of Pennsylvania, Philadelphia, PA 19104-6396, USA} \\[.3cm]
$\,^4$ {Center for Applied Mathematics and Theoretical Physics,\\ University of Maribor, Maribor, Slovenia}\\[.3cm]

vbraun$@$map.cc, cvetic$@$physics.upenn.edu, donagi$@$math.upenn.edu,  
maximilian$_-$poretschkin$@$yahoo.de

\end{center}

\vskip 0.0cm

\begin{center} {\bf ABSTRACT } \end{center}
We provide the first explicit example of  Type IIB string theory compactification  on a  globally defined  Calabi-Yau threefold with torsion which results in a four-dimensional effective theory with a non-Abelian discrete gauge symmetry.  Our example is based on a particular Calabi-Yau manifold, the quotient of a product of three elliptic curves by a fixed point free action of $\IZ_2 \times \IZ_2$.  
Its cohomology contains torsion classes in various degrees. The main technical novelty is in determining the multiplicative structure of the (torsion part of) the cohomology ring, and in particular showing that the cup product of second cohomology torsion elements goes non-trivially to the fourth cohomology. This specifies a non-Abelian, Heisenberg-type discrete symmetry group of the four-dimensional theory. 
\end{titlepage}

\section{Introduction}

Discrete symmetries are an integral part of particle physics; they play a key r\^ole in the Standard Model physics, such as those to prevent nucleon and lepton number violating processes,  with R-symmetry being the most prominent one, as well as examples of 
Abelian and non-Abelian discrete symmetries to address the flavor hierarchy.
 
 In string theory,  discrete symmetries are expected to be gauge symmetries. In recent years there has been a flurry of activities in studies of discrete symmetries in string and F-theory compactification, focusing on their geometric origin.
In F-theory  the primary focus was on the origin of Abelian discrete symmetries, which arise  from  Calabi-Yau geometries which are genus-one fibrations without section, in contrast to elliptic fibrations with sections. A
natural object associated with  these compactifications is  the Tate-Shafarevich
(TS) group 
%
%
%
which is a discrete Abelian group that organizes inequivalent genus-one geometries which share the same associated Jacobian fibration. The TS group specifies the Abelian discrete gauge symmetry of the F-theory compactification.
The study of F-theory compactifications  with discrete gauge symmetries $\mathbb{Z}_n$ was initiated  in \cite{Braun:2014oya} and followed-up in  \cite{Morrison:2014era, Anderson:2014yva, Klevers:2014bqa, Garcia-Etxebarria:2014qua,Mayrhofer:2014haa,Mayrhofer:2014laa,Cvetic:2015moa,Cvetic:2016}.  
Most of the past works  primarily  focused on $\mathbb{Z}_2$  gauge symmetry. However,  new insights into aspects of the TS group, and its relations to M-theory vacua in the case of $\mathbb{Z}_3$ were addressed in \cite{Cvetic:2015moa}. Furthermore, there has been progress in elucidating the origin of Abelian discrete symmetries via F-theory/Heterotic duality \cite{Cvetic:2016}.

On the other hand, the origin of non-Abelian discrete gauge symmetries in string theory is less understood. An important progress in this direction was made \cite{Uranga} in the context of Type II
string theory compactification to four-dimensions, building on  earlier works \cite{CIM,BISU,ISU} that study Abelian discrete gauge symmetries in Type II string theory compactification on Calabi-Yau threefolds with torsion. In the context of Type IIB string theory,  a non-Abelian Heisenberg-type  discrete symmetry  is realized \cite{Uranga}  on a  Calabi-Yau threefold with torsion whose second cohomology torsion  elements have a non-trivial cup product into fourth cohomology torsion ones. 
This specific approach therefore  requires the study of  Calabi-Yau threefolds with torsion by  determining  torsion cohomology groups and their cup products,  which is technically  challenging.

The purpose of this paper is to provide a first explicit construction of a
Type IIB string theory compactification  on a Calabi-Yau manifold that exhibits a Heisenberg-type discrete symmetry.\footnote{Non-Abelian discrete symmetries  of Type IIB string theory on $AdS_5\times S^5/\Z_3$  were studied in \cite{Gukov:1998kn} from the perspective of AdS/CFT correspondence.} 
For that purpose we choose an example of a Calabi-Yau threefold $X_6$ with torsion. 
This particular Calabi-Yau  threefold occurs in the classification of \cite{DW} as the first example of the {\em free} quotient  
of a torus $T^6$ by a finite group action. It had occurred previously, in various contexts, in \cite{Ig,Ue,O}. 
It is the quotient of a product of three elliptic curves by a fixed point free action of $\IZ_2 \times \IZ_2$ which we describe explicitly in \eqref{eq:action}.
The integer cohomology of our Calabi-Yau manifold $X_6$ is given in \eqref{eq:Xcohomology}.
To the best of our knowledge, this is the first example of a self-mirror
Calabi-Yau threefold where the two ({a} priori independent) torsion
groups $\text{Torsion}\, (H^2(X_6,\Z)) = \text{Torsion}\, (H^5(X_6,\Z)) = \Z_4^2 \oplus \Z_2^3$ and
$\text{Torsion}\, (H^3(X_6,\Z)) = \text{Torsion}\, (H^4(X_6,\Z))=\Z_2^3$ actually differ and are both non-zero.

This  example allows us to carry out the challenging computation of the cup products for torsion coholomology elements explicitly, which in turn determine the Heisenberg discrete symmetry of the four-dimensional theory.
The key  strategy is to relate $X_6$ to a lower dimensional (real) submanifold $Y_0$. Recalling that $X_6$ is the quotient of a six-torus by a group $G=\IZ_2 \times \IZ_2$, we take $Y_0$ to be the quotient of a 4-dimensional subtorus of $X_6$ that happens to be invariant under $G$. 
It comes with an inclusion map $i: Y_0 \to  X_6$ and a projection map going the other way, hence 
the restriction $i^*: H^*(X_6,\IZ) \to H^*(Y_0,\IZ)$ is surjective, 
and in fact it exhibits $H^*(Y_0,\IZ)$ as a direct summand of $H^*(X_6,\IZ)$.
The cup product  calculation on the {\it smaller} $Y_0$ can be carried out explicitly.
This could be done by hand using elementary topology. Instead, we obtain it as part of a computational scheme that gives us many additional useful facts about the manifolds involved,
including the integer cohomology of $X$, given in \eqref{eq:Xcohomology}.

The cup product on $Y_0$ turns out to have the property we want: 
there are torsion classes in $H^2(Y_0,\IZ)$ whose product does not vanish in $H^4(Y_0,\IZ)$.
Even though we do not fully calculate the muliplicative structure of the cohomlogy ring of X, 
the fact that $H^*(Y_0,\IZ)$ is a direct summand of $H^*(X_6,\IZ)$
plus knowing the multiplicative structure of the cohomlogy ring of $Y_0$ suffices to allow us to conclude that there are torsion classes in $H^2(X_6,\IZ)$ 
whose product does not vanish in $H^4(X_6,\IZ)$, as claimed. 
In particular, Type IIB compactification on such a Calabi-Yau manifold
leads to a four-dimensional  theory with a Heisenberg-type discrete
symmetry group.\footnote{For a related work in the context of
  F-theory, see \cite{Grimm:2016} where the origin of the Heisenberg-type
  discrete symmetry does not seem to have a weak coupling limit to
  Type IIB string theory. In the context of Heterotic string
  compactifications some aspects of torsion homologies were studied,
  e.g., in \cite{Braun:2007} for the Schoen manifold.} 

In section \ref{Non-Abelian}  we summarize the results of \cite{Uranga} regarding Type IIB string theory compactifications leading to  four-dimensionall theories with  Heisenberg-type discrete symmetries. Our Calabi-Yau manifold is described in section \ref{CY}. The rest of the paper is devoted to the calculation of the torsion homologies and cup products for this manifold. In section \ref{submanifolds} we determine the torsion cohomology groups for the Calabi-Yau manifold and some of its submanifolds,  and in section \ref{Cup} the specific results for the cup products are obtained.
The outlook is given in section \ref{Outlook}.
 
\section{Non-Abelian discrete symmetries in Type IIB} \label{Non-Abelian}

In this section, we review the construction of non-Abelian discrete symmetries arising in Type IIB compactifications on a Calabi-Yau manifold $X_6$.  The analysis follows closely that of \cite{Uranga}.
This latter  analysis is closely related to the  study \cite{CIM} of Abelian discrete symmetries in Type IIB compactifications. 

In general, the homology of a threefold $X_6$ has two independent torsion sub-groups given by

\ban
\text{Torsion} \, (H_1(X_6, \mathbb{Z})) &\simeq &\text{Torsion} \, (H_4(X_6, \mathbb{Z}))\, , \nn\\\text{Torsion}\, ( H_2(X_6, \mathbb{Z})) &\simeq& \text{Torsion}\, ( H_3(X_6, \mathbb{Z}))\, ,
\ean
associated with the torsion one-cycles (and  Poincar{\' e}  dual torsion four-cycles) and torsion 
two-cycles  (and  Poincar{\' e}  dual torsion three-cycles), respectively.
%
%
%

In the following, we shall first restrict our discussion to the case that 
\ban
\text{Torsion} \, (H_1(X_6, \mathbb{Z}))& \simeq& \text{Torsion}\, ( H_4(X_6, \mathbb{Z})) = \mathbb{Z}_k\, , \nn\\
\text{Torsion}\, ( H_2(X_6, \mathbb{Z}) )&\simeq &\text{Torsion} \, (H_3(X_6, \mathbb{Z}))= \mathbb{Z}_{k'} \, .
\ean
The Poincar\'e dual cohomology groups which are needed for the dimensional reduction of Ramond-Ramond fields  are accordingly given as
\ban
\text{Torsion} \, (H^5(X_6, \mathbb{Z})) &\simeq &\text{Torsion}\, ( H^2(X_6, \mathbb{Z})) = \mathbb{Z}_k\, , \nn\\ \text{Torsion} \, (H^4(X_6, \mathbb{Z})) &\simeq &\text{Torsion}\, ( H^3(X_6, \mathbb{Z})) = \mathbb{Z}_{k'} \, .
\ean
Let  $\rho_2$,  $\beta_3$, ${\tilde \omega}_4$, and $\zeta_5$  represent the generators of the torsion cohomologies \\ $\text{Torsion}\, ( H^2(X_6, \mathbb{Z}))$,  $ \text{Torsion}\, (H^3(X_6, \mathbb{Z}))$,  $ \text{Torsion} \, (H^4(X_6, \mathbb{Z}))$ and $ \text{Torsion}\, ( H^5(X_6, \mathbb{Z})) $, respectively. 
They satisfy the following relations 
\begin{eqnarray}
d \gamma_1 &=& k \rho_2, \qquad d \tilde{\rho}_4 = k \zeta_5, \nn \\
d \alpha_3 &=& k' \tilde{\omega}_4, \qquad d \omega_2 = k' \beta_3 \, , \label{nonco}
\end{eqnarray}
where $\gamma_1$, $\omega_2$, $\alpha_3$ and ${\tilde \rho}_4$ are  non-closed  one-, two-, three-  and four-forms on $X_6$, respectively, and they satisfy: 
\begin{equation}
\int_{X_6}\gamma_1\wedge \zeta_{5}=\int_{X_6}\rho_2\wedge \tilde\rho_{4}=\int_{X_6}\alpha_3\wedge \beta_{3}=\int_{X_6}\omega_{2}\wedge\tilde\omega_4=1\, .\label{poin}
\end{equation}
Here $k^{-1}$ and $k'^{-1}$ are the torsion linking numbers between dual  torsion $p$- and $(5-p)$-cycles ($p=1,3$).   Note, eqs.(\ref{nonco}) and  (\ref{poin}) can be obtained from  expressions  that determine torsion linking numbers, c.f., appendix C of \cite{CIM}.

The cup-product of two torsion classes is again a torsion class. Thus the product $\rho_2 \wedge \rho_2$ is some multiple of the generator ${\tilde \omega}_4$ of $\text{Torsion}\, (H^4(X_6;\mathbb{Z})$: 
\begin{equation}
\rho_2\wedge\rho_2=M\, \tilde\omega_4\, , \quad \quad M\in \IZ\, . \label{origin}
\end{equation}
The coefficient $M $ is an invariant of the manifold $X_6$. Sometimes it vanishes, and sometimes it does not. In this work we describe an example where it is non zero.
 By employing (\ref{nonco}) this cup-product integrates to   $\rho_2\wedge \gamma_1=M'\, \alpha_3 $, where  $M' \in \IZ$ and $k \, M=k'\, M'$.

These torsion subgroups give {a} priori rise to three non-commuting discrete cyclic groups in the effective four-dimensional Type IIB action. This can be seen from the following Kaluza-Klein reduction Ansatz for the  Type IIB  closed string sector Ramond-Ramond (RR)  and Neveu-Schwarz-Neveu-Schwarz (NSNS) two-form fields  $C_2, B_2$, respectively  and RR four-form field $C_4$: 
\begin{eqnarray}
B_2 &=& b^1 \wedge \rho_2 + A^1 \wedge \gamma_1\, \\
C_2 &=& b^2 \wedge \rho_2 + A^2 \wedge \gamma_1\, \\
C_4 &=& b^3 \tilde{\omega}_4 + A^3 \wedge \alpha_3 + V^3 \wedge \beta_3 + c_2 \wedge \omega_2\, , \label{kk}
\end{eqnarray}
where $b^i$  and $A^i$ ($i=1,2,3$)  are the three axions and three $U(1)$ one-form gauge potentials, respectively.  
(One-form  potential $V^3$ and two-form potential $c_2$, are not independent fields, due to the self-duality of the five-form field strength $F_5$  in Type IIB supergravity.)  

These Ans\"atze ensure that the Kaluza-Klein reduction of Type IIB supergravity results in an effective four-dimensional  field theory with  three massive $U(1)$ one-form gauge  potentials $A^i$ ($=i=1,2,3$).  (For further details, see  section 4 of \cite{Uranga}.) E.g., for $B_2$ one obtains:
\begin{eqnarray}
&&\int_{M_{10}=M_4\times X_6} d B_2 \wedge * d B_2 \longrightarrow\\
&& \int_{M_4} (d b^1 - k A^1) \wedge * (d b^1 - k A^1) \int_{X_6} \rho_2 \wedge * \rho_2 + \int_{M_4} (d  A^1) \wedge * (d A^1) \int_{X_6} \gamma_1 \wedge * \gamma_1\, , \nonumber
\end{eqnarray}
which results in  a St\"uckelberg mass term for $A^1$.
The St\"uckelberg mass contributions for all three gauge fields $A^1$, $A^2$, $A^3$ in the effective four-dimensional action is of the following schematic form:
\begin{equation}
{\cal L}\supset 
{\cal G}_{ij} \, \eta_\mu^i \eta^{\mu\, j}\, ,
\label{lag}
\end{equation}
where 
\begin{eqnarray}
\eta_\mu^i&=&  \partial_\mu b^i- k\, A_\mu^i \, , \quad \quad i=1,2\, , \nonumber\\
\eta_\mu^3&=& \partial_\mu b^3- k' A_\mu ^{3}-Mb^2(\partial_\mu b^1-k\,A^1_\mu)\, .
\end{eqnarray}
This four-dimensional action is therefore  invariant under the following non-commuting discrete gauge transformations:
\begin{eqnarray}
A^i_\mu \to A^i_\mu + \partial_\mu\lambda^i \ , && \quad   b^i\to b^1+k\lambda^i \, , \quad\quad i=1,2\, , \nonumber\\
A^3_\mu \to A^3_\mu + \partial_\mu\lambda^3 +M' k\lambda^2A^1_\mu+M'b^1\partial_\mu \lambda^2\, ,  && \quad
b^3 \to b^3 + Mk b^1\lambda^2+k'\lambda^3\, ,\label{gauget}
\end{eqnarray}
where $M\in\IZ$, $M'\in \IZ$ and  $kM=k'M'$. This corresponds to a set of non-commuting $\mathbb{Z}_{k}$, $\mathbb{Z}_k$, $\mathbb{Z}_{k'}$  factors as long as $M\ne 0$,  resulting in a non-Abelian discrete gauge symmetry of the four-dimensional action,  specified by $k$, $k'$  and $M$. 

Altogether  there are three generators 
$T_1, T_2, T_3$ associated with the discrete symmetry groups $\mathbb{Z}_k, \mathbb{Z}_k, \mathbb{Z}_{k'}$, respectively. 
The important fact to note is that these generators $T_1, T_2, T_3$ do not commute, provided that there is a non-trivial cup-product (\ref{origin}).

These discrete gauge symmetries of the effective four-dimensional action lead to the following discrete symmetry operations on a four-dimensional state $\psi (x)$, with charges $(q_1,q_2,q_3)$ under $(\mathbb{Z}_k, \, \mathbb{Z}_k, \mathbb{Z}_{k'})$ :
\begin{eqnarray}
\tilde{T}_1: \psi(x) &\longrightarrow& e^{2\pi i k^{-1}q_1} \psi(x) \nn \\
\tilde{T}_2: \psi(x) &\longrightarrow& e^{2\pi i k^{-1}q_2} \mathcal{U}\psi(x) \nn \\
\tilde{T}_3: \psi(x) &\longrightarrow& e^{2\pi i k'^{-1}q_3} \psi(x) \, ,
\end{eqnarray}
where  the charge redefinition matrix ${\cal U}$  is of the form:
\begin{equation}
\begin{pmatrix} q_1 \\ q_2 \\ q_3 \end{pmatrix} \mapsto \begin{pmatrix} 1 & 0 & M' \\ 0 & 1 & 0 \\ 0 & 0 & 1 \end{pmatrix}  \begin{pmatrix} q_1 \\ q_2 \\ q_3 \end{pmatrix} \, , \qquad k M = k' M'\, .
\end{equation}
Thus, one observes that 
\be
\tilde{T}_1 \tilde{T}_2 = \tilde{T}_3^M \tilde{T}_2 \tilde{T}_1\, ,
\ee
resulting in a non-commuting discrete gauge symmetry, a Heisenberg discrete symmetry  group  $(\mathbb{Z}_k\times\mathbb{Z}_{k'})\rtimes \mathbb{Z}_k$, specified by $k$, $k'$ and $M$.  In special cases, say, 
 when  $k=k'$, $M=1$ the non-Abelian discrete gauge symmetry is given by $(\mathbb{Z}_k\times\mathbb{Z}_k)\rtimes \mathbb{Z}_k$ and further specializations of $k$ reduce  to, e.g.,  $Dih_4$ for $k=2$ and $\Delta(27)$ for $k=3$.

\subsection{Generalizations}

It is straightforward  to generalize this analysis to the case  when the second torsion cohomologies have multiple discrete factors.  (For further details, see \cite{Grimm:2016}, section 2.)

Let us focus on  the following specific examples, which shall be relevant for the rest of our analysis.   The torsion cohomologies are chosen to be:
\ban
\text{Torsion}\, (H^5(X_6, \mathbb{Z}) )&\simeq &\text{Torsion}\, (H^2(X_6,\IZ))=\mathbb{Z}_{k_1}\times \mathbb{Z}_{k_2}\, , \nn\\ \text{Torsion}\, (H^4(X_6,\IZ))&\simeq &\text{Torsion} \, (H^3(X_6, \mathbb{Z})) =\mathbb{Z}_{k_3}\, , 
\ean
 and  the nontrivial cup product of the generators $\rho_2^{(i)} \in \mathbb{Z}_{k_i}$ ($i=1,2$)  is of the following form:
\begin{equation}
\rho_2^{(1)}\wedge \rho_2^{(2)}=M{\tilde \omega}_4\, ,
\end{equation}
where ${\tilde \omega}_4$ is the  generator of  $\text{Torsion}\, (H^4(X_6,\IZ))=\mathbb{Z}_{k_3}$ .

These generators satisfy the following relations\footnote{For the sake of simplicity, we chose specialized relations; for the analysis of more general cases, see \cite{Grimm:2016}.}:\begin{eqnarray}
d \gamma_1^{(i)} &=& k_i \rho_2^{(i)}, \qquad d \tilde{\rho}_{4(i)} = k_i \zeta_{5(i)}\,,\quad \quad  i=1,2\, , \nn \\
d \alpha_3 &=& k_3 \tilde{\omega}_4, \qquad d \omega_2 = k_3 \beta_3 \, , \label{nonc}
\end{eqnarray}
where $\beta_3$, and $\zeta_{5 (i)}$ ($i=1,2$)  represent the generators of the torsion cohomologies  $ \text{Torsion}\, (H^4(X_6, \mathbb{Z}))$ and $ \text{Torsion} \, (H^5(X_6, \mathbb{Z})) $, respectively, and $\gamma_{1(i)} $, $\omega_2$, $\alpha_3$ and ${\tilde \rho}_{4(i)}$ are  non-closed  one-, two-, three-  and four-forms that satisfy: 
\begin{equation}
\int_{X_6}\gamma_1^{(i)} \wedge \zeta_{5(j)}=\int_{X_6}\rho_2^{(i)}\wedge \tilde\rho_{4 (j)}=\delta^i_{j}\, , \quad \int_{X_6}\alpha_3\wedge \beta_{3}=\int_{X_6}\omega_{2}\wedge\tilde\omega_4=1\, .
\end{equation}
 Kaluza-Klein Ans\"atze  for $B_2$, $C_2$ and $C_4$ gauge potentials, parallel those of (\ref{kk}):
 \begin{eqnarray}
B_2 &=& b^{1}_{(i)} \wedge \rho_2^{(i)} + A^{1}_{(i)} \wedge \gamma_{1}^{(i)}\, \\
C_2 &=& b^{2}_{(i)} \wedge \rho_2 ^{(i)}+ A^{2}_{(i)} \wedge \gamma_{1}^{(i)}\, \\
C_4 &=& b^3 \tilde{\omega}_4 + A^3 \wedge \alpha_3 + V^3 \wedge \beta_3 + c_2 \wedge \omega_2\, , \label{kkp}
\end{eqnarray}
In the four-dimensional effective action there are  five  massive U(1) gauge fields $A^{1}_{(i)}$, $A^{2}_{(i)}$ and $A^3$, and five associated  axions $b^{1}_{(i)}$,  $b^{2}_{(i)}$,  and  $b^3$, respectively.  (Again,  $V^3$ and $c_2$  are not independent fields, due to the self-duality of  $F_5$.)

The St{\"u}ckelberg mass contributions for  to the effective action again takes the schematic form:
\begin{equation}
{\cal L}\supset 
{\cal G}_{IJ*} \, \eta_\mu^{I} \eta^{\mu\, J*}\,  ,
\label{lagp}
\end{equation}
where  $\eta_\mu^I$, complexified four-vectors, which take the following form:
\begin{eqnarray}
\eta_{\mu(i)}&=&  \partial_\mu b^2_{(i)}-\tau \partial_\mu b^1_{(i)}+ k_i \left(A^2_{\mu(i)}-\tau A^1_{\mu(i)}\right) \, , \quad \quad i=1,2\, , \nonumber\\
\eta_\mu^3&=& \partial_\mu b^3+ k_3A_\mu ^{3}-M\left(b^2_{(1)}-\tau b^1_{(1)}\right)k_2\,A^1_{\mu(2)}\, .
\end{eqnarray} 
and $\tau=C_0+{\rm i} {\rm e}^{-\phi}$ denotes the complexified string
coupling of Type IIB string theory.  This structure results in   the
discrete gauge invariance of the effective four-dimensional action, 
which corresponds to the Heisenberg discrete symmetry specified by $k_1$, $k_2$, $k_3$ and $M$.
For further details see \cite{Uranga} section 2 and \cite{Grimm:2016}, section 3.
%
%
%
%
%

Thus, in order to determine the Heisenberg  discrete group of  Type IIB string compactifications on a Calabi-Yau threefold with torsion,
the plan is to identify second cohomology torsion  classes and to determine their non-trivial cup products.  As explained in the introduction, we proceed to relate the Calabi-Yau
threefold $X_6$ with torsion to a simpler space $Y_0$, a submanifold, 
where the cup product is under control.
In particular, we exhibit a torsion class $t$ in the second cohomology $H^2(X_6,\IZ)$ 
whose restriction to $Y_0$  is non-zero and squares to a non-
zero class on the auxiliary $Y_0$. Functoriality of this cup product then fixes
the rest. 
In this paper we apply this strategy  the example of the Calabi-Yau threefold $X_6$, defined in the section below, and  explicit calculations are derived in the subsequent two sections \ref{submanifolds} and \ref{Cup}.

\section{The Calabi-Yau Manifold}\label{CY}

Our Calabi-Yau threefold $X$\footnote{For simplicity, in the rest of the paper we shall omit the subscript $6$ for a Calabi-Yau threefold, i.e. $X_6\to X$.}
will be the quotient of a six-torus (in fact the product of three elliptic curves) by a finite group action. The first and best known example of such a quotient was studied by Vafa and Witten in \cite{VW}. 
Let
\begin{math}
  E_i = \C / (\Z + \tau_i \Z)
  \owns z_i
\end{math}, be three elliptic curves, $i = 0, 1, 2$. Their product
admits an action of the group $G=\Z_2\times\Z_2$, 
generated by the transformations;
\begin{equation}
  \label{eq:action}
  \begin{aligned}
    g_1^0: (z_0, z_1, z_2) \mapsto&\, 
    \big(z_0 ,\ -z_1,\ -z_2\big),
    \\
    g_2^0: (z_0, z_1, z_2) \mapsto&\, 
    \big(-z_0,\ z_1,\ -z_2 \big).
  \end{aligned}
\end{equation}
Vafa and Witten then consider a crepant resolution of this quotient. 

Various other quotients, by different group actions, were considered by Oguiso and Sakurai \cite{O}
in the process of studying the collection of all Calabi-Yau manifolds with an infinite fundamental group. 
They obtained a partial classification, which has very recently been completed in \cite{Ka}.
All actions of the basic group $G=\Z_2\times\Z_2$ and of all its Abelian extensions  were classified in \cite{DW}. (Those actions of Abelian extensions that specialize to the Vafa-Witten action on $G$ had been classified earlier, in \cite {DF}.) It turns out that, up to obvious equivalences, there are four such actions of $G$, cf. \cite{DW}, Lemma 1.2.2. They all have the same linearization, so they differ only in the shifts. Exactly one of these $G$ actions is fixed-point free:
the fixed-point free $G=\Z_2\times\Z_2$  action is
generated by the transformations:
\begin{equation}
  \label{eq:actionp}
  \begin{aligned}
    g_1: (z_0, z_1, z_2) \mapsto&\, 
    \big(z_0 + \tfrac{1}{2},\ -z_1,\ -z_2\big),
    \\
    g_2: (z_0, z_1, z_2) \mapsto&\, 
    \big(-z_0,\ z_1 + \tfrac{1}{2},\ -z_2 + \tfrac{1}{2}\big).
  \end{aligned}
\end{equation}
This modifies the Vafa-Witten action by adding some non-trivial shifts. It is these shifts that make the action fixed-point free, and therefore the quotient: 
\begin{equation} \label{X}
  X = (E_0 \times E_1 \times E_2) / G
\end{equation}
is a manifold, with no need for a resolution. This particular quotient is described in example 2.17 of \cite{O}, where it is attributed to Igusa \cite{Ig}, page 678, and to Ueno \cite{Ue}, Example [16.16].
Each of $g_1, g_2$ and $g_3:=g_1\circ g_2$ preserves 
the constant holomorphic $(3,0)$-form on $E_0 \times E_1 \times E_2$,
so this form descends to a nowhere vanishing holomorphic $(3,0)$-form on $X$.
On the other hand, the $g_i$ project out any
constant one-forms, so the quotient $X$ of \eqref{X}
is a proper Calabi-Yau threefold in the sense that compactification
preserves only the minimal amount of supersymmetry. We note that its
holonomy group is $G$, and its fundamental group is easily seen to be 
the semidirect product
\begin{equation}
  \pi_1(X) = \Z^6 \rtimes G,
\end{equation}
cf. \cite{DW}, section 1.5 and Table 1 on page 10. Finally, the only invariant $(1,1)$-forms are $dz_i\wedge dz_i$,
$i=0,1,2$, leading to the Hodge diamond
\begin{equation}
  h^{pq}\big(X\big)
  =
  \vcenter{\xymatrix@!0@=6mm@ur{
      1 &  0 &  0 & 1 \\
      0 &  3 &  3 & 0 \\
      0 &  3 &  3 & 0 \\
      1 &  0 &  0 & 1 
    }}
  .
\end{equation}

For further details  see \cite{DW}. In this work we will analyze the cohomology ring of this Calabi-Yau manifold $X$ of \eqref{X}. In \cite{DW} it was noted (Lemma 1.7.1) that this is one of four topologically inequivalent free quotients (the others are by various abelian extensions of $G$). All are Calabi-Yau manifolds with Hodge numbers $(3,3)$, and it seems plausible that similar calculations can be carried out for each of these fixed-point free actions. We will not pursue these other manifolds in the present work. 

\section{Submanifolds}\label{submanifolds}

Since the group action eq.~\eqref{eq:actionp} only ever changes the
imaginary part of the coordinates $z_i$ by a sign, there are a number
of $G$-invariant (real) submanifolds of the product, hence submanifolds of the quotient $X$,
obtained by setting the imaginary part to zero. On the
covering space, these are sub-tori of $E_0\times E_1 \times
E_2$. After dividing out the group action, we obtain the special
Langrangian 3-manifold
\begin{equation}
  \begin{aligned}
    Y \hookrightarrow&\, X,
    &
    (
    x_0,~
    x_1,~
    x_2
    )
    \mapsto
    (
    x_0,~
    x_1,~
    x_2
    ),
  \end{aligned}
\end{equation}
three 4-dimensional submanifolds
\begin{equation}
  \begin{aligned}
    Y_0 \hookrightarrow&\, X,
    &
    (
    x_0,~
    x_1,~
    x_2,~
    y_0
    )
    \mapsto
    (
    x_0+\tau_0 y_0,~ 
    x_1,~
    x_2
    ),
    \\
    Y_1 \hookrightarrow&\, X,
    &
    (
    x_0,~
    x_1,~
    x_2,~
    y_1
    )
    \mapsto
    (
    x_0,~
    x_1+\tau_1 y_1,~
    x_2
    ),
    \\
    Y_2 \hookrightarrow&\, X,
    &
    (
    x_0,~
    x_1,~ 
    x_2,~ 
    y_2
    )
    \mapsto
    (
    x_0,~
    x_1,~
    x_2+\tau_2 y_2
    ),
    \\
  \end{aligned}
\end{equation}
and three 5-dimensional submanifolds
\begin{equation}
  \begin{aligned}
    Y_{01} \hookrightarrow&\, X,
    &
    (
    x_0,~
    x_1,~ 
    x_2,~
    y_0,~ 
    y_1
    )
    \mapsto
    (
    x_0+\tau y_0,~ 
    x_1+\tau y_1,~
    x_2
    ),
    \\
    Y_{02} \hookrightarrow&\, X,
    &
    (
    x_0,~
    x_1,~
    x_2,~
    y_0,~ 
    y_2
    )
    \mapsto
    (
    x_0+\tau y_0,~ 
    x_1,~
    x_2+\tau y_2
    ),
    \\
    Y_{12} \hookrightarrow&\, X,
    &
    (
    x_0,~
    x_1,~
    x_2,~
    y_1,~
    y_2
    )
    \mapsto
    (
    x_0,~ 
    x_1+\tau y_1,~
    x_2+\tau y_2
    ).
    \\
  \end{aligned}
\end{equation}
In addition to the submanifold embeddings, we note that there are also
projection maps $X\rightarrow Y_{ij}$, $X\rightarrow Y_i$, and
$X\rightarrow Y$ by ignoring the imaginary part of one, two, and all
three complex coordinates. Therefore, these submanifolds are all
retractions and the relative cohomology long exact sequences split into
\begin{equation}
  \label{eq:split}
  \begin{aligned}
    H^*(X,\Z) \simeq&\,
    H^*(Y,\Z) \oplus H^*(X, Y,\Z)
    \\
    \simeq&\,
    H^*(Y_i,\Z) \oplus H^*(X, Y_i,\Z),
    &&
    0 \leq i < 3,
    \\
    \simeq&\,
    H^*(Y_{ij},\Z) \oplus H^*(X, Y_{ij},\Z),
    &\quad&
    0 \leq i < j < 3.
  \end{aligned}
\end{equation}
In the remainder of this section we now discuss the integral cohomology
of these submanifolds.

\subsection{The Special Lagrangian Submanifold \boldmath$Y$}
\label{sec:Y}

Again, all constant 2-forms on the covering space torus are projected
out by the $G$-action. Hence, this is a rational homology sphere. Its
fundamental group and Abelianization is
\begin{equation}
  \pi_1(Y) = \Z^3 \rtimes G
  ,\quad
  H_1(Y) =  \pi_1 / [\pi_1, \pi_1] = \Z_4 \oplus \Z_4.
\end{equation}
To summarize, the integral cohomology is
\begin{equation}
  H^d(Y, \Z) =
  \begin{cases}
    \Z & d=3 \\
    \Z_4\oplus \Z_4 & d=2 \\
    0 & d=1 \\
    \Z & d=0 \\
  \end{cases}
\end{equation}
and, by degree count, there can be no non-trivial cup products.

To get some explicit handle on the torsion cohomology generators, let
us write down cochains in a cellular model. Our model of choice is
going to be cubical cells with identifications. First, by cubical
cells we mean (hyper-)cubes in some $\R^n$ whose vertices are
integral, edges are parallel to the axes, and side lengths are either
zero or one only. We write such a cube as
\begin{equation}
  [a_{0,0}, a_{0,1}] \times [a_{1,0}, a_{1,1}] 
  \times \cdots \times 
  [a_{n-1, 0}, a_{n-1, 1}]
\end{equation}
subject to the constraints
\begin{itemize}
\item $a_{k,l} \in \Z$ for all $0\leq k < n$, $l\in\{0,1\}$,
\item $a_{k,0} = a_{k,1}$ or $a_{k,0} + 1 = a_{k,1}$ for all $0\leq k < n$.
\end{itemize}
To efficiently represent tori and quotients thereof, we also allow
identifications of cubical cells, possibly with the opposite
orientation. Hence, we will use equivalence classes of oriented
cubical cells as our basic {\it cell}, which we will write as
\begin{equation}
  [a_{0,0}, a_{0,1}] \times [a_{1,0}, a_{1,1}] 
  \times \cdots \times 
  [a_{n-1, 0}, a_{n-1, 1}]
  \big/ \sim
\end{equation}
and always refer to as cubical cells / cell complex in the following.

Although cubical cells are clearly convenient for describing tori, one
might wonder whether the existence of a cubical cell complex imposes any
constraint on a manifold. In fact, for topological manifolds we
now known~\cite{Manolescu} that 
non-triangulable manifolds, that is, manifolds not homeomorphic to any
simplicial complex, exist in each dimension $\geq 4$. 
However, for smooth manifolds such as ours, the situation is
notably different, and triangulations always exist, by an old theorem
of Whitehead. Finally, we note that the existence of a triangulation
and of a cubical cell decomposition are equivalent. In one direction this
equivalence is clear: the barycentric subdivision of any polyhedron
(in particular, of a cube) yields a decomposition into
simplices. Conversely, consider a $d$-simplex and its barycentric
subdivision into $(d+1)!$ top-dimensional simplices. For each of the
original $d+1$ vertices, note that the $d!$ adjacent sub-simplices fit
together into a $d$-dimensional cube. In fact, this is the dual
polyhedral decomposition of the simplex, and it defines a canonical
cubical decomposition of a triangulation. Therefore, a smooth manifold
always has a decomposition into a cubical cell complex.

Coming back to $Y$, we start with eight maximal cells\footnote{Before
  any identifications, the cubical complex consists of 8 three-cubes,
  36 squares, 54 line segments, and 27 vertices.}
\begin{equation}
  \begin{aligned}
  &[0,1]\times[0,1]\times[0,1]
  ,\quad&
  &[0,1]\times[0,1]\times[1,2]
  ,~\\
  &[0,1]\times[1,2]\times[0,1]
  ,\quad&
  &[0,1]\times[1,2]\times[1,2]
  ,~\\
  &[1,2]\times[0,1]\times[0,1]
  ,\quad&
  &[1,2]\times[0,1]\times[1,2]
  ,~\\
  &[1,2]\times[1,2]\times[0,1]
  ,\quad&
  &[1,2]\times[1,2]\times[1,2].
  \end{aligned}
\end{equation}
By identifying opposite sides, this is a $3$-torus with real
coordinates
\begin{equation}
  (\xi_0, \xi_1, \xi_2) = 
  (2 x_0, 2 x_1, 2 x_2) \in 
  (\R/2\Z)^3,
\end{equation}
and, using these coordinates, the $G$ action becomes (compare \cite{DW}, page 6 and Lemma 1.2.2):
\begin{equation}
  \label{eq:Yaction}
  \begin{aligned}
    g_1: (\xi_0, \xi_1, \xi_2) \mapsto&\, 
    \big(\xi_0 + 1,\ -\xi_1, -\xi_2\big),
    \\
    g_2: (\xi_0, \xi_1, \xi_2) \mapsto&\, 
    \big(-\xi_0,\ \xi_1 + 1,\ -\xi_2 + 1\big).
  \end{aligned}
\end{equation}
The equivalence classes of cells under both the group action and
identification of opposite sides in listed in \autoref{tab:Ycells}.
\begin{table}[p]
  \vspace{-2cm}
  \hspace{-2cm}
  \setlength{\jot}{1pt}
  \begin{tabular}{cc|cc}
    dim & cell & same orientation & opposite orientation \\
    \hline
    $3$ & 
    $[0,1] \times [0,1] \times [0,1]/\sim$ &
    \scriptsize
    \begin{math}
      \begin{aligned}
        {}\{
        &[0,1] \times [0,1] \times [0,1],~
         [1,2] \times [1,2] \times [1,2], \\
        &[0,1] \times [0,1] \times [1,2],~
         [1,2] \times [1,2] \times [0,1] \}
      \end{aligned}
    \end{math}
    &
    \scriptsize
    \begin{math}
      \{\}
    \end{math}
    \\\hline
    $3$ & 
    $[0,1] \times [1,2] \times [0,1]/\sim$ &
    \scriptsize
    \begin{math}
      \begin{aligned}
        {}\{
        &[1,2] \times [0,1] \times [1,2],~
         [0,1] \times [1,2] \times [1,2], \\
        &[0,1] \times [1,2] \times [0,1],~
         [1,2] \times [0,1] \times [0,1] \}
      \end{aligned}
    \end{math}
    &
    \scriptsize
    \begin{math}
      \{\}
    \end{math}
    \\\hline
    $2$ & 
    $[0,0] \times [0,1] \times [0,1]/\sim$ &
    \scriptsize
    \begin{math}
      \begin{aligned}
        {}\{
        &[0,0] \times [0,1] \times [0,1], \\
        &[1,1] \times [1,2] \times [1,2], \\
        &[2,2] \times [0,1] \times [0,1]
        \}
      \end{aligned}
    \end{math}
    &
    \scriptsize
    \begin{math}
      \begin{aligned}
        {}\{
        &[0,0] \times [1,2] \times [0,1], \\
        &[1,1] \times [0,1] \times [1,2], \\
        &[2,2] \times [1,2] \times [0,1]
        \}
      \end{aligned}
    \end{math}
    \\\hline
    $2$ & 
    $[0,0] \times [0,1] \times [1,2]/\sim$ &
    \scriptsize
    \begin{math}
      \begin{aligned}
        {}\{
        &[0,0] \times [0,1] \times [1,2], \\
        &[1,1] \times [1,2] \times [0,1], \\
        &[2,2] \times [0,1] \times [1,2]
        \}
      \end{aligned}
    \end{math}
    &
    \scriptsize
    \begin{math}
      \begin{aligned}
        {}\{
        &[0,0] \times [1,2] \times [1,2], \\
        &[1,1] \times [0,1] \times [0,1], \\
        &[2,2] \times [1,2] \times [1,2]
        \}
      \end{aligned}
    \end{math}
    \\\hline
    $2$ & 
    $[0,1] \times [0,0] \times [0,1]/\sim$ &
    \scriptsize
    \begin{math}
      \begin{aligned}
        {}\{
        &[0,1] \times [0,0] \times [0,1], \\
        &[0,1] \times [2,2] \times [0,1], \\
        &[1,2] \times [1,1] \times [0,1]
        \}
      \end{aligned}
    \end{math}
    &
    \scriptsize
    \begin{math}
      \begin{aligned}
        {}\{
        &[1,2] \times [0,0] \times [1,2], \\
        &[1,2] \times [2,2] \times [1,2], \\
        &[0,1] \times [1,1] \times [1,2]
        \}
      \end{aligned}
    \end{math}
    \\\hline
    $2$ & 
    $[0,1] \times [0,0] \times [1,2]/\sim$ &
    \scriptsize
    \begin{math}
      \begin{aligned}
        {}\{
        &[0,1] \times [0,0] \times [1,2], \\
        &[0,1] \times [2,2] \times [1,2], \\
        &[1,2] \times [1,1] \times [1,2]
        \}
      \end{aligned}
    \end{math}
    &
    \scriptsize
    \begin{math}
      \begin{aligned}
        {}\{
        &[1,2] \times [0,0] \times [0,1], \\
        &[1,2] \times [2,2] \times [0,1], \\
        &[0,1] \times [1,1] \times [0,1] 
        \}
      \end{aligned}
    \end{math}
    \\\hline
    $2$ & 
    $[0,1] \times [0,1] \times [0,0]/\sim$ &
    \scriptsize
    \begin{math}
      \begin{aligned}
        {}\{
        &[0,1] \times [0,1] \times [1,1], \\
        &[0,1] \times [0,1] \times [2,2], \\
        &[0,1] \times [0,1] \times [0,0]
        \}
      \end{aligned}
    \end{math}
    &
    \scriptsize
    \begin{math}
      \begin{aligned}
        {}\{
        &[1,2] \times [1,2] \times [1,1], \\
        &[1,2] \times [1,2] \times [2,2], \\
        &[1,2] \times [1,2] \times [0,0]
        \}
      \end{aligned}
    \end{math}
    \\\hline
    $2$ & 
    $[0,1] \times [1,2] \times [0,0]/\sim$ &
    \scriptsize
    \begin{math}
      \begin{aligned}
        {}\{
        &[0,1] \times [1,2] \times [1,1], \\
        &[0,1] \times [1,2] \times [2,2], \\
        &[0,1] \times [1,2] \times [0,0]
        \}
      \end{aligned}
    \end{math}
    &
    \scriptsize
    \begin{math}
      \begin{aligned}
        {}\{
        &[1,2] \times [0,1] \times [1,1], \\
        &[1,2] \times [0,1] \times [2,2], \\
        &[1,2] \times [0,1] \times [0,0]
        \}
      \end{aligned}
    \end{math}
    \\\hline
    $1$ & 
    $[0,0] \times [0,0] \times [0,1]/\sim$ &
    \scriptsize
    \begin{math}
      \begin{aligned}
        {}\{
        &[2,2] \times [2,2] \times [0,1],~
         [2,2] \times [0,0] \times [0,1], \\
        &[1,1] \times [1,1] \times [1,2],~
         [0,0] \times [2,2] \times [0,1], \\
        &[0,0] \times [0,0] \times [0,1]
        \}
      \end{aligned}
    \end{math}
    &
    \scriptsize
    \begin{math}
      \begin{aligned}
        {}\{
        &[2,2] \times [1,1] \times [0,1], \\
        &[1,1] \times [2,2] \times [1,2], \\
        &[1,1] \times [0,0] \times [1,2], \\
        &[0,0] \times [1,1] \times [0,1]
        \}
      \end{aligned}
    \end{math}
    \\\hline
    $1$ & 
    $[0,0] \times [0,0] \times [1,2]/\sim$ &
    \scriptsize
    \begin{math}
      \begin{aligned}
        {}\{
        &[0,0] \times [0,0] \times [1,2],~
         [2,2] \times [2,2] \times [1,2], \\
        &[0,0] \times [2,2] \times [1,2],~
         [2,2] \times [0,0] \times [1,2], \\
        &[1,1] \times [1,1] \times [0,1]
        \}
      \end{aligned}
    \end{math}
    &
    \scriptsize
    \begin{math}
      \begin{aligned}
        {}\{
        &[1,1] \times [2,2] \times [0,1],~
         [1,1] \times [0,0] \times [0,1], \\
        &[2,2] \times [1,1] \times [1,2],~
         [0,0] \times [1,1] \times [1,2]
        \}
      \end{aligned}
    \end{math}
    \\\hline
    $1$ & 
    $[0,0] \times [0,1] \times [0,0]/\sim$ &
    \scriptsize
    \begin{math}
      \begin{aligned}
        {}\{
        &[0,0] \times [0,1] \times [2,2],~
         [0,0] \times [0,1] \times [0,0], \\
        &[2,2] \times [1,2] \times [1,1],~
         [2,2] \times [0,1] \times [0,0], \\
        &[0,0] \times [1,2] \times [1,1],~
         [2,2] \times [0,1] \times [2,2]
        \}
      \end{aligned}
    \end{math}
    &
    \scriptsize
    \begin{math}
      \begin{aligned}
        {}\{
        &[1,1] \times [1,2] \times [2,2], \\
        &[1,1] \times [1,2] \times [0,0], \\
        &[1,1] \times [0,1] \times [1,1]
        \}
      \end{aligned}
    \end{math}
    \\\hline
    $1$ & 
    $[0,0] \times [0,1] \times [1,1]/\sim$ &
    \scriptsize
    \begin{math}
      \begin{aligned}
        {}\{
        &[0,0] \times [1,2] \times [2,2],~
         [0,0] \times [1,2] \times [0,0], \\
        &[2,2] \times [0,1] \times [1,1],~
         [2,2] \times [1,2] \times [0,0], \\
        &[0,0] \times [0,1] \times [1,1],~
         [2,2] \times [1,2] \times [2,2]
        \}
      \end{aligned}
    \end{math}
    &
    \scriptsize
    \begin{math}
      \begin{aligned}
        {}\{
        &[1,1] \times [0,1] \times [2,2], \\
        &[1,1] \times [0,1] \times [0,0], \\
        &[1,1] \times [1,2] \times [1,1]
        \}
      \end{aligned}
    \end{math}
    \\\hline
    $1$ & 
    $[0,1] \times [0,0] \times [0,0]/\sim$ &
    \scriptsize
    \begin{math}
      \begin{aligned}
        {}\{
        &[1,2] \times [2,2] \times [2,2],~
         [0,1] \times [2,2] \times [2,2], \\
        &[1,2] \times [2,2] \times [0,0],~
         [0,1] \times [0,0] \times [0,0], \\
        &[1,2] \times [0,0] \times [2,2],~
         [0,1] \times [2,2] \times [0,0], \\
        &[0,1] \times [0,0] \times [2,2],~
         [1,2] \times [0,0] \times [0,0]
        \}
      \end{aligned}
    \end{math}
    &
    \scriptsize
    \begin{math}
      \begin{aligned}
        {}\{
        &
        \}
      \end{aligned}
    \end{math}
    \\\hline
    $1$ & 
    $[0,1] \times [0,0] \times [1,1]/\sim$ &
    \scriptsize
    \begin{math}
      \begin{aligned}
        {}\{
        &[0,1] \times [0,0] \times [1,1],~
         [1,2] \times [2,2] \times [1,1], \\
        &[1,2] \times [0,0] \times [1,1],~
         [0,1] \times [2,2] \times [1,1]
        \}
      \end{aligned}
    \end{math}
    &
    \scriptsize
    \begin{math}
      \begin{aligned}
        {}\{
        &[0,1] \times [1,1] \times [0,0],~
         [1,2] \times [1,1] \times [0,0], \\
        &[1,2] \times [1,1] \times [2,2],~
         [0,1] \times [1,1] \times [2,2]
        \}
      \end{aligned}
    \end{math}
    \\\hline
    $0$ & 
    $[0,0] \times [0,0] \times [0,0]/\sim$ &
    \scriptsize
    \begin{math}
      \begin{aligned}
        {}\{
        15~\text{points}
        \}
      \end{aligned}
    \end{math}
    &
    \scriptsize
    \begin{math}
      \{\}
    \end{math}
    \\\hline
    $0$ & 
    $[0,0] \times [0,0] \times [1,1]/\sim$ &
    \scriptsize
    \begin{math}
      \begin{aligned}
        {}\{
        12~\text{points}
        \}
      \end{aligned}
    \end{math}
    &
    \scriptsize
    \begin{math}
      \{\}
    \end{math}
  \end{tabular}
  \caption{Cubical cell complex for $Y$. For each equivalence class, 
    the cubical cells with the same and opposite orientation
    are shown.}
  \label{tab:Ycells}
\end{table}
Using this notation, the two generators of $H^2(Y,\Z)=\Z_4\oplus \Z_4$
can be written as cochains
\begin{equation}
  \label{eq:YZ4Z4}
  \begin{split}
    \hat{c}_1 =&\,
    \chi\big([0,1] \times [0,1] \times [0,0] / \sim\big),
    \\
    \hat{c}_2 =&\,
    \chi\big([0,0] \times [0,1] \times [0,1] / \sim\big) +
    \chi\big([0,1] \times [0,0] \times [1,2] / \sim\big),
  \end{split}
\end{equation}
where $\chi(c)$ denotes the cochain dual to the cell $c$, that is, the
cochain that evaluates to one on $c$ and to zero on all other cells.

\subsection{The Four-Dimensional Submanifold \boldmath$Y_0$}
\label{sec:Y0}

Note that $Y_0$ is not orientable, so its top cohomology group is
$\Z_2$. The fundamental group and Abelianization of $Y_0$ is
\begin{equation}
  \pi_1(Y_0) = \Z^4 \rtimes G
  ,\quad
  H_1(Y_0) =  \pi_1 / [\pi_1, \pi_1] = 
  \Z_2 \oplus \Z_4 \oplus \Z_4 \oplus \Z.
\end{equation}
Finally, its degree-3 cohomology is not going to be relevant for
even-degree cup products in the following, but can easily be
determined numerically from the cell complex structure. To summarize,
the integral cohomology is
\begin{equation}
  \label{eq:Y0cohomology}
  H^d(Y_0, \Z) =
  \begin{cases}
    \Z_2 & d=4 \\
    \Z^2 & d=3 \\
    \Z_2 \oplus \Z_4\oplus \Z_4 \oplus \Z& d=2 \\
    0 & d=1 \\
    \Z & d=0. \\
  \end{cases}
\end{equation}
We observe that the degrees are such that there can be a non-trivial
cup product $H^2 \times H^2 \to H^4$ involving torsion cohomology
classes, which we will investigate in \autoref{sec:Y0cup}.

The cubical complex for $Y_0$ is very similar to \autoref{tab:Ycells}, the
only change is that we add a factor $\times [0,1]$ for the $y_0$
coordinate, that is, use coordinates
\begin{equation}
  (\xi_0, \xi_1, \xi_2, \eta_0) = 
  (2x_0, 2x_1, 2x_2, y_0) \in
  (\R/2\Z)^3 \times (\R/\Z).
\end{equation}
Note that the group action 
\begin{equation}
  \label{eq:Yactionp}
  \begin{aligned}
    g_1: (\xi_0, \xi_1, \xi_2, \eta_0) \mapsto&\, 
    \big(\xi_0 + 1,\ -\xi_1, -\xi_2, \eta_0\big),
    \\
    g_2: (\xi_0, \xi_1, \xi_2, \eta_0) \mapsto&\, 
    \big(-\xi_0,\ \xi_1 + 1,\ -\xi_2 + 1, -\eta_0\big).
  \end{aligned}
\end{equation}
never shifts $\eta_0$, which is why we do not need any subdivision in
the cubical complex in that direction. After identifying opposing
sides and $G$-images, we can again write down explicit cochains for
the cohomology classes of interest. The generator of
$\Z_2\subset H^2(Y_0,\Z)$ can be chosen to be the 2-cochain
\begin{equation}
  \label{eq:Y0Z2}
  \begin{split}
    c_0 =&\, 
    \chi\big([0,0] \times [0,0] \times [0,1] \times [0,1] / \sim\big) 
    \\
    &\quad
    -\chi\big([0,0] \times [0,0] \times [1,2] \times [0,1] / \sim\big).
  \end{split}  
\end{equation}
By the retraction property, the 4-torsion part
$\Z_4\times\Z_4\subset H^2(Y_0, \Z)$ is necessarily the pullback of
$H^2(Y,\Z)$. Hence the generators are
\begin{equation}
  \label{eq:Y0Z4Z4}
  \begin{split}
    c_1 =&\,
    \chi\big([0,1] \times [0,1] \times [0,0] \times [0,0] / \sim\big),
    \\
    c_2 =&\,
    \chi\big([0,0] \times [0,1] \times [0,1] \times [0,0] / \sim\big) 
    \\&\quad
    +\chi\big([0,1] \times [0,0] \times [1,2] \times [0,0] / \sim\big),
  \end{split}
\end{equation}
see eq.~\eqref{eq:YZ4Z4}. Finally, a free $\Z\subset H^2(Y_0,\Z)$ is generated by
\begin{equation}
  \label{eq:Y0Z}
  \begin{split}
    c_3 =&\, 
    \chi\big([0,1] \times [0,0] \times [0,0] \times [0,1] / \sim\big) 
    \\
    &\quad
    + \chi\big([0,1] \times [0,0] \times [1,1] \times [0,1] / \sim\big)
  \end{split}  
\end{equation}

\subsection{Cohomology of the Calabi-Yau Manifold}

We know already by eq.~\eqref{eq:split} that $H^*(Y_0,\Z)$ is
a direct summand of the cohomology of the Calabi-Yau manifold $X$. As far
as cup products are concerned, this is all that we will be using in
the following. However, for completeness let us note that the entire
cohomology group can be computed numerically from the cubical cell
complex, and the result is
\begin{equation}
  \label{eq:Xcohomology}
  H^d(X, \Z) =
  \begin{cases}
    \Z & d=6 \\
    \Z_4^2 \oplus \Z_2^3 & d=5 \\
    \Z^3 \oplus \Z_2^3  & d=4 \\
    \Z^8 \oplus \Z_2^3 & d=3 \\
    \Z^3 \oplus \Z_4^2 \oplus \Z_2^3 & d=2 \\
    0 & d=1 \\
    \Z & d=0. \\
  \end{cases}
\end{equation}
To the best of our knowledge, this is the first example of a self-mirror
Calabi-Yau threefold where the two ({a} priori independent) torsion
groups $H^2(X,\Z)_\tors = H^5(X,\Z)$ and
$H^3(X,\Z)_\tors = H^4(X,\Z)_\tors$ actually differ.

\section{Cup Product}\label{Cup}

\subsection{Cup Product on \boldmath$Y_0$}
\label{sec:Y0cup}

For orientable manifolds, the cup product is dual to the cap
(intersection) product. Now $Y_0$ is not orientable, so Poincar\'e
duality does not hold over $\Z$. However, any manifold is
$\Z_2$-orientable which is sufficient for our purposes since the
codomain of the cup product
\begin{equation}
  \cup: H^2(Y_0,\Z) \times H^2(Y_0,\Z) \to H^4(Y_0,\Z) =\Z_2
\end{equation}
is two-torsion anyways. Furthermore, all relevant intersections turn
out to be transversal, which lets us read off the cup product from the
cochain representatives in eqns.~\eqref{eq:Y0Z2}, \eqref{eq:YZ4Z4},
and \eqref{eq:Y0Z}. The result is that
\begin{equation}
  c_0 \cup c_1 = c_2 \cup c_3 \not= 0
\end{equation}
and all other products vanish.

\subsection{Naturality and the Calabi Yau Manifold}

Recall that the cup product is natural, that is, the diagram
\begin{equation}
  \xymatrix{
    H^*(X,\Z) \times H^*(X,\Z)
    \ar[d]^{f^*}
    \ar[r]^-\cup & 
    H^*(X,\Z)
    \ar[d]^{f^*} \\
    H^*(Y_0,\Z) \times H^*(Y_0,\Z)
    \ar[r]^-\cup & 
    H^*(Y_0,\Z) 
  }
\end{equation}
commutes for any map $f:Y_0\to X$. When applied to our embedding map
$i:Y_0 \hookrightarrow X$, we note that $i^*$ is surjective by
eq.~\eqref{eq:split}. In particular, there are elements $\bar{c}_i \in
H^2(X,\Z)$ such that $i^*(\bar{c}_i) = c_i$ are our generators of
$H^2(Y_0,\Z)$, $i=0,1,2,3$. Their cup products
\begin{equation}
  \bar{c}_0 \cap \bar{c}_1
  ,~
  \bar{c}_2 \cap \bar{c}_3
  \in H^4(X,\Z)
\end{equation}
must be non-trivial cohomology classes because 
\begin{equation}
  i^*(\bar{c}_0 \cap \bar{c}_1) = 
  i^*(\bar{c}_2 \cap \bar{c}_3) \not= 0
  \in H^4(Y_0,\Z)=\Z_2
\end{equation}

To summarize, the resulting non-commuting discrete gauge symmetries  of four dimensional theory are  associated with $\mathbb{Z}_2\times \mathbb{Z}_4$ sectors of second torsion cohomology and  a $\mathbb{Z}_2$ sector of  the fourth torsion cohomology,   resulting in the Heisenberg group determined by $k_1=2$, $k_2=4$,  $k_3=2$ and  $M=1$.\footnote{Note that a cup product with the free sector of $H^2$ does not result in a non-Abelian discrete symmetry.}
  
\section{Outlook} \label{Outlook}

In this paper we provided the first explicit example of Type IIB string theory compactification on  a Calabi-Yau manifold, which leads to  a non-Abelian discrete gauge symmetry in four-dimensions. The compactification is based on the  Calabi-Yau  threefold whose torsion cohomology structure results  in a non-trivial cup product of the second cohomology  torsion class elements, thus resulting in a non-Abelian gauge symmetry associated with a Heisenberg-type discrete group.

We should, however, point out that this is a very specific Type IIB compactification on a smooth Calabi-Yau manifold.  Generalizations to Calabi-Yau orientifolds, which would also allow for introduction of non-Abelian continuous symmetries and  chiral matter,  are expected to be straightforward, with generators  $\rho_2$  (${\tilde \omega}_4$) of the second (fourth) cohomology torsion classes  odd (even) under the orientifold action\footnote{Note however, that issues  related to the manifestly supersymmetric  four-dimensional action were raised in \cite{Grimm:2016}.}.  Further studies of such compactifications with a  full fledged particle physics content would be an important future research direction.

We should also note that  these  constructions result  in specific non-Abelian discrete gauge symmetries, namely,  Heisenberg-type  ones. Furthermore, specific results apply  here only to four-dimensional field theories, and not to six-dimensional ones,  as $K3$ Calabi-Yau twofolds have no torsion classes.  Therefore an important generalization
to other Calabi-Yau manifolds, both twofolds and threefolds,  where the non-Abelian discrete symmetries could arise, e.g., from generalized isometries of the compactification manifold, awaits further studies. 

Last, but not least, an important direction of this program is to extend studies of non-Abelian discrete gauge symmetries  to F-theory,  and to shed light on the  geometric origins of non-Abelian discrete gauge symmetries there. Techniques spelled out in  this paper may  open the door for a systematic construction of F-theory  compactifications on  elliptically fibered Calabi-Yau fourfolds  with a structure of torsion cohomologies  \cite{Grimm:2016} that  would result in four-dimensional non-Abelian discrete symmetries.

\vskip 0.8cm

\noindent {\large \bf Acknowledgements}
\vskip 0.5cm
\noindent We would like to thank Jonathan Block for help with an earlier version of this work, Ling Lin for numerous discussions and  a collaboration on related topics, and Angel Uranga for useful communications.
The  research of Mirjam Cveti\v c  is
supported in part by the DOE Grant Award de-sc0013528 (M.C., M.P.), the Fay R. and
Eugene  L.  Langberg  Endowed  Chair  (M.C.) and  the  Slovenian  Research  Agency (ARRS) (M.C.).
During the preparation of this work Ron Donagi was supported in part
by NSF grant DMS 1603526 and by Simons Foundation grant \# 390287.

\end{document}